\documentclass[conference]{IEEEtran}
\IEEEoverridecommandlockouts

\usepackage{enumitem}
\usepackage{cite}
\usepackage{amsmath,amssymb,amsfonts}
\usepackage{algorithmic}
\usepackage{graphicx}
\usepackage{textcomp}
\usepackage{xcolor}
\usepackage{hyperref}
\usepackage{booktabs}
\usepackage{balance}
\usepackage{makecell}
\usepackage{array}
\usepackage{tikz}
\usepackage{pifont}
\newcommand{\xmark}{\ding{55}}%
\def\BibTeX{{\rm B\kern-.05em{\sc i\kern-.025em b}\kern-.08em
    T\kern-.1667em\lower.7ex\hbox{E}\kern-.125emX}}
\usepackage{xargs}                      
\begin{document}

\title{ \centering A Survey of Automatic Generation of  \\ Attack Trees
and Attack Graphs
}

\author{\IEEEauthorblockN{Alyzia-Maria Konsta}
\IEEEauthorblockA{
\textit{DTU Compute}\\
Kongens Lyngby, Denmark \\
akon@dtu.dk}
\and
\IEEEauthorblockN{Beatrice Spiga}
\IEEEauthorblockA{
\textit{DTU Compute}\\
Kongens Lyngby, Denmark \\
s225953@student.dtu.dk}
\and
    
\IEEEauthorblockN{Alberto Lluch Lafuente}
\IEEEauthorblockA{
\textit{DTU Compute}\\
Kongens Lyngby, Denmark \\
albl@dtu.dk
}

\and

\IEEEauthorblockN{Nicola Dragoni}
\IEEEauthorblockA{
\textit{DTU Compute}\\
Kongens Lyngby, Denmark \\
ndra@dtu.dk
}

}

\maketitle

\begin{abstract}
Graphical security models constitute a well-known, user-friendly way to represent the security of a system. These classes of models are used by security experts to identify vulnerabilities and assess the security of a system. The manual construction of these models can be tedious, especially for large enterprises. Consequently, the research community is trying to address this issue by proposing methods for the automatic generation of such models. In this work, we present a survey illustrating the current status of the automatic generation of two kinds of graphical security models -Attack Trees and Attack Graphs. The goal of this survey is to present the current methodologies used in the field, compare them, and present the challenges and future directions for the research community.
\end{abstract}


\section{Introduction}
During the last few decades, the use of electronic devices has spread significantly. Companies and individuals are using technology for both personal and professional reasons. In the last decade, with the use of IoT devices in all aspects of our everyday lives, a huge amount of personal and sensitive data has been stored or processed on computer networks. These kinds of devices are limited in resources, and in some cases, the cryptographic algorithms are not applicable. Along with the extended use of electronic devices, the attempts by malicious users to steal this data have increased. As a consequence, the research community has focused on finding ways to protect our data from malicious actors.

One well-known solution for assessing the security of a system is the use of graphical security models. These models represent security scenarios and help security experts identify the system's weaknesses. Their graphical representation constitutes a user-friendly way to analyze the security of the system. In the scope of this paper, we are going to examine only two kinds of these graphical representations: Attack Trees~\cite{schneier1999attack} and Attack Graphs~\cite{Phillips}. 
For a comprehensive overview of all the available formalisms, we refer the reader to~\cite{KORDY}. Given our focus, we will use the term ``Attack Models'' to refer to the class of graphical security models constituted by both Attack Trees and Attack Graphs.

Currently, security experts manually produce graphical security models. This procedure, especially for such large systems used nowadays, can be tedious, error-prone, and non-exhaustive. Consequently, the need for automatic procedures arose, and the research community has focused on addressing this issue. Especially in the last decade, researchers have been studying different methods to automatically produce graphical security models. In this work, we present an exhaustive survey of the automatic generation of Attack Models. We provide an overview of the field and identify current challenges and future research opportunities. To our knowledge, this is the first survey paper focusing only on the automatic generation of Attack Models.

\paragraph*{Scope}

Among all known graphical models of security, we restrict our attention to Attack Trees and Attack Graphs, due to their popularity in the scientific community. The popularity of Attack Trees over alternative models is claimed in various papers, starting with the survey of Kordy et al.~\mbox{\cite{KORDY}}. The survey compares Attack Trees to other graphical models of security, namely Attack Graphs, models based on Petri Nets such as ``privilege graphs'' and models based on UML such as ``mal-activity diagrams''.
The popularity of Attack Graphs and Attack Trees is also evident in bibliographic databases. For instance, the query ``attack tree'' in Google Scholar yields 7,620 papers, ``attack graph'' yields 9,210 papers, ``privilege graphs'' yields 174 papers; and ``mal-activity diagrams'' yields 186 papers. The prevalence of Attack Trees and Attack Graphs is more evident if the query is combined with ``threat modeling''.

\paragraph*{Research Questions}
The main research questions in our work are
\begin{itemize}
\item \textbf{RQ1:} What kinds of techniques have been proposed for automatically generating Attack Models?
\item 
\textbf{RQ2:} What are the characteristics of the proposed approaches in terms of formal rigor, empirical evidence, reproducibility, scalability, etc.?
\item \textbf{RQ3:} What are the challenges and opportunities in this area of research?
\end{itemize}

We believe that addressing these questions will help identify current limitations on existing techniques and gaps in the literature in order to provide future directions to the research community.
Moreover, addressing these questions will help readers navigate through the literature on automated generation of attack models in a way that makes it easy to see how techniques relate to each other, which tools are available, and ultimately how to identify which method suits their case best.

\paragraph*{Contributions}
The main contributions are
\begin{enumerate}
\item A structured literature review of existing approaches to automated generation of attack models, which addresses \textbf{RQ1} and \textbf{RQ2} by discussing all existing papers under a common categorization framework. As we shall see, one of the main outcomes of our paper is that the existing approaches generate attack models from (combinations of) \emph{models} of the system under study, \emph{analysis results} of the system under study, or \emph{well-known vulnerability patterns}. We examine all papers through these lenses and look into key aspects of those as demanded by \textbf{RQ2}.
\item An identification of limitations and future directions for the research community in order to reduce the gap between research developments and practitioners' needs, therefore answering \textbf{RQ3}. One of our main outcomes is a general lack of reproducibility aspects and open tool availability in the proposed approaches. 
\end{enumerate}

The paper is structured as follows: Section~\ref{sec:4} presents the related work, Section~\ref{sec:2} presents the main concepts of Attack Models, Section~\ref{sec:cat} presents the categorization framework used to classify the papers under study, Section~\ref{sec:3} presents the research method used to structure our research, Section~\ref{sec:class} presents the classification of the papers, Section~\ref{sec:lim} discusses limitations, challenges and future research directions, and Section~\ref{sec:con} concludes the paper.

\section{Related Work}\label{sec:4}

\begin{table*}[!t]\centering\begingroup\fontsize{10pt}{10pt}\selectfont
  \caption{Overview of the papers included in the Related Work}
  \label{tab:rw}
  \centering
\begin{tabular}{||c | c | c | c | c ||} 
 \hline
 Paper & After 2015 & \makecell[c]{Automatic\\ Generation}  & \makecell[c]{Techniques' \\ Classification} & Challenges \\ [0.5ex] 
 \hline\hline
 \cite{KORDY} &  &  & \checkmark & \checkmark\\ 
 \hline
 \cite{lallie2020review} & \checkmark &  &  & \checkmark\\
 \hline
 \cite{widel2019beyond} & \checkmark & \checkmark &  & \checkmark\\
 \hline
 Our Work & \checkmark & \checkmark & \checkmark & \checkmark\\
 \hline
\end{tabular}
\endgroup
\end{table*}

In this section, we discuss other surveys that consider the issue of the automatic generation of Attack Models. We chose these works since they provide surveys on Attack Models, and some of them also consider their automatic generation. We also include a table to summarize the main characteristics of the papers included in the Related Work section. In Table~\ref{tab:rw} four different characteristics are presented. We denote the symbol \checkmark when the referred paper fulfills the corresponding characteristic. The first column includes a reference to the paper under examination; the second column, ``After 2015'' refers to the papers that have been published after 2015. This is an important characteristic since more than half of the papers included in our survey were published after 2015. The third column, ``Automatic Generation'' refers to the papers that study the automatic generation of Attack Models, the fourth column refers to whether the paper presents a classification for the papers included in their survey based on the techniques used in each paper; and the last column indicates whether the paper indicates the challenges identified in the field.
The survey papers that we have identified are the following:

Kordy et al.~\cite{KORDY} presents a very comprehensive survey of DAG-based graphical models for security. Their work summarizes the state of the art of the existing methodologies, they compare them and classify them based on their formalism. Although this is a very extensive survey, they do not focus on the automated generation of these models. Since it does not cover works published after 2015, their work does not cover several papers on Attack Model generation, that are instead covered in our work.

Lallie et al.~\cite{lallie2020review} examine the effectiveness of the visual representation of Attack Models. They analyze how these structures represent the cyberattacks. They conclude that there is no standard method to represent the Attack Models and that the research community should turn their attention to standardizing the representation. Although this paper is a great contribution, it does not examine the issue of the automatic generation of these structures.

Wojciech et al.~\cite{widel2019beyond} provides a survey of the application of formal methods on Attack Trees, their semi-automated or automated generation, and their quantitative analysis. Although they consider automatic generation, their research is not only focused on that. Also, they do not consider the automatic generation of Attack Graphs. Moreover, this paper considered only formal methods approaches and hence does not cover some of the papers we have studied, for example, machine learning approaches.

In our work, we discuss the different methodologies applied by the research community for automating the generation of Attack Models. To our knowledge, this is the first survey focusing on the automatic generation of Attack Models. To be more specific, the automatic generation of Attack Trees is only studied in~\cite{widel2019beyond}, where all the papers referring to Attack Graphs are not considered-~\cite{ou2005mulval, Sheyner, sheyner2003tools, Ghazo, ibrahim2019automatic, Swiler, Tippenhauer, Phillips, koo2022attack, bezawada2019agbuilder, Ingols}, we consider all of the above papers. Also, the following papers that generate Attack Tree are not considered-~\cite{Bryans, Kumar, Pinchinat_2, jhawar2018semi, Hong} in~\cite{widel2019beyond}, on the other hand, we consider all the papers mentioned above. Overall, our survey covers 15 papers that are not covered in any of the existing surveys.

\section{Background}\label{sec:2}

Before starting our survey we recall here the basic concepts of Attack Models. 

\subsection{Attack Trees}

\begin{figure}[]
\centering
\includegraphics[width=\columnwidth]{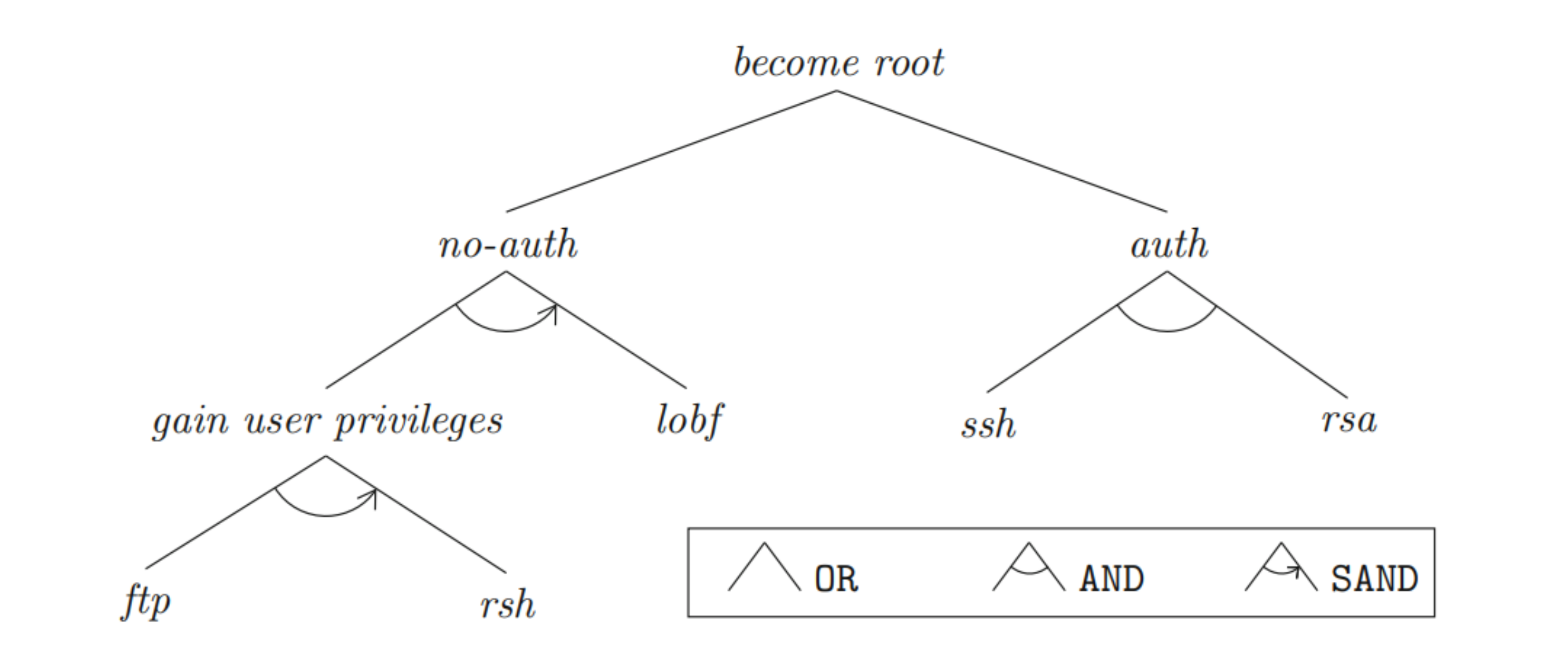}
\caption{Attack Tree from~\cite{Jhawar}}
\label{fig:attack_tree}
\end{figure}

An Attack Tree is a graphical representation of potential attacks on the system's security in the form of a tree. Initially introduced by Schneiner~\cite{schneier1999attack}, this type of representation enables developers to identify the vulnerabilities of the system and facilitates the implementation of countermeasures. They model attack scenarios presented in a hierarchical way, with each labeled node corresponding to a sub-goal of the attacker and the \textbf{root} being the main one—the global goal of the attack. The rest of the labeled nodes can be either \textbf{children of a node}, a refinement of the parent's goal into subsidiary goals, or \textbf{leaf nodes}, representing attacks that cannot be further refined, also called basic actions.
An Attack Tree is a 3-tuple $(N, \rightarrow, n_0)$, where $N$ is a finite set of nodes, $\rightarrow$ is a finite acyclic relation of type $\rightarrow \subseteq N \times M(N)$, where $M(N)$ is the multi-set of $N$, and $n_0$ is the root node, such that every node in $N$ is reachable from $n_0$~\cite{mauw2006foundations}.

The basic formal model of Attack Trees incorporates two types of refinements: \texttt{OR} and \texttt{AND}. \texttt{OR} nodes represent disjunction (choice), where the parent node's goal is achieved when at least one of the children's sub-goals is achieved, \texttt{AND} nodes represent conjunction (aggregation), thus requiring all children's sub-goals to be fulfilled. Several variants of Attack Trees have been proposed. For example, the sequential conjunction refinement \texttt{SAND}, is similar to \texttt{AND} but requires a specific sequential realization of the children~\cite{widel2019beyond,mauw2006foundations}. One example of an Attack Tree is illustrated in Figure~\ref{fig:attack_tree}. In summary, there are two separate ways for the attacker to accomplish their goal, namely, by becoming a root: with or without authentication. The two refined authentication options are ssh and rsa, and both must be fulfilled since they are under an \texttt{AND} operator. On the other hand, without authentication, the user must first gain user privileges, fulfill actions ftp and then rsh. Following the acquisition of privileges, lobf should be achieved.
The tree notation is very appealing and convenient for a threat analysis process since it can include multiple attacks derived from physical, technical, and even human vulnerabilities~\cite{widel2019beyond}.

\subsection{Attack Graphs}

\begin{figure}[]
\centering
\includegraphics[width=\columnwidth]{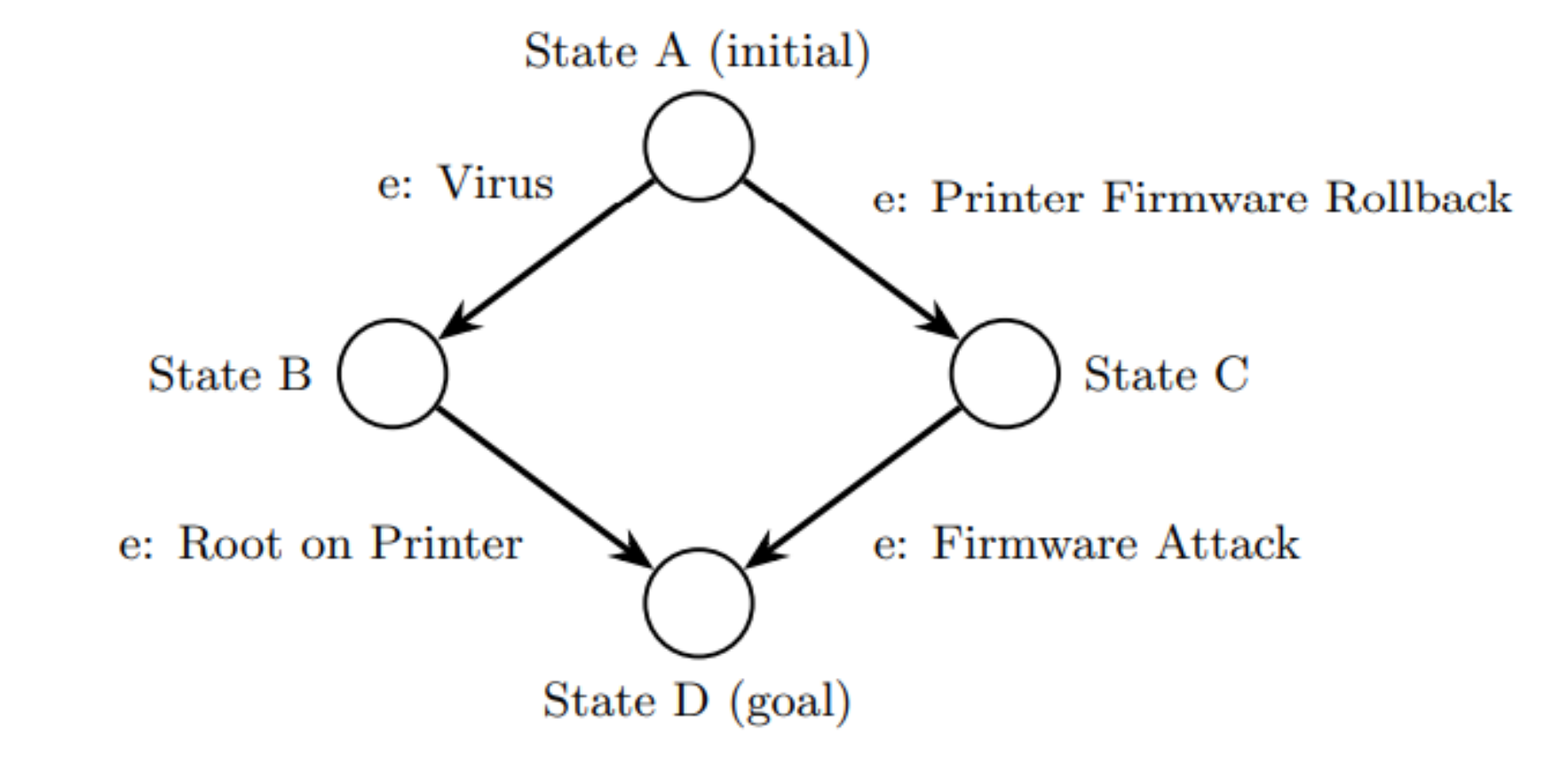}
\caption{Attack Graph from~\cite{cook}
}
\label{fig:attack_gr}
\end{figure}

Attack graphs are graphical representations that depict all the possible paths an attacker can follow to achieve their malicious goal. They are frequently used to represent complex attacks that have multiple paths and goals~\cite{lallie2020review}. Attack Graphs directly show the existence of vulnerabilities in the system and how attackers can exploit these vulnerabilities to implement an effective attack~\cite{Zhong}. 

An Attack Graph or AG is a tuple \begin{math}G = (S, \rightarrow, S_0, S_s)\end{math}, where S is a set of states, \begin{math}\rightarrow \subseteq S \times S\end{math} is a transition relation, \begin{math}S_0 \subseteq S\end{math} is a set of initial states, and \begin{math}S_s \subseteq S\end{math} is a set of success states~\cite{Sheyner}. Intuitively, \begin{math}S_s\end{math} denotes the set of states where the intruder has achieved their goals. Unless stated otherwise, we assume that the transition relation \begin{math}\tau\end{math} is total. We define an execution fragment as a finite sequence of states \begin{math}s_0, s_1, ..., s_n\end{math} such that \begin{math}(s_i, s_i+1) \in \tau\end{math} for all \begin{math}0 \leq i \leq n\end{math}. An execution fragment with \begin{math}s_0 \in S_0\end{math} is an execution, and an execution whose final state is in \begin{math}S_s\end{math} is an attack, i.e., the execution corresponds to
a sequence of atomic attacks leading to the intruder’s goal~\cite{Sheyner}. 

One example of an Attack Graph can be seen in Figure~\ref{fig:attack_gr}. The example represents the Attack Graph of a simple network. There is one switch connected to the internet on one side and to a workstation and a printer on the other end. The Attack Graph represents different states of the network, and the edges are exploits that an attacker can use~\cite{cook}. In the specific example, the virus can exploit a particular vulnerability in the network in order to transition to State B, where it can exploit another vulnerability to gain root access.

\subsection{Attack Trees vs Attack Graphs} 

A graph can be seen as a set of objects connected by a set of edges. A graph can be either directed or undirected. A directed graph has edges that represent a specific direction; one node is set to be the origin and the other to be the destination. The edges in an undirected graph do not specify which node is the origin and which is the end point; they just denote a two-way connection.

A tree is a special case of a graph, more specifically a Directed-Acyclic Graph (DAG). A tree is represented by three different types of nodes: the root node, which does not have any parents; the internal nodes, which have both a parent and at least one child; and the leaf nodes, which do not have children. So, the tree represents a hierarchy between the nodes.

Taking the above into consideration, we can conclude that the basic structure of Attack Trees and Attack Graphs is different. The Attack Graph can represent more than one goal node, or in some cases, even circles, if an attacker is trying the same unsuccessful action multiple times. On the other hand, the Attack Trees only represent one goal node and constitute an acyclic graph by definition. We can also observe the different visual representations in Figure~\ref{fig:attack_tree} and Figure~\ref{fig:attack_gr}.
Furthermore, in Attack Graphs, the event flow is represented top-down, while in the majority of Attack Trees, this is depicted in a bottom-up way~\cite{lallie2020review}. Moreover, in the same sense, Attack Trees generally use vertices to represent exploits and not preconditions, while preconditions are assumed to have been met in the transition from one exploit to the next. Attack graphs represent both~\cite{lallie2020review}. Essentially, Attack Models have a graph-based structure. The main differences are: how the event flow is depicted, the representation of full and partial attacks, and the representation of preconditions.

\section{Categorisation Framwework}\label{sec:cat}

\begin{figure*}[t]
\centering
\includegraphics[width=0.9\textwidth]{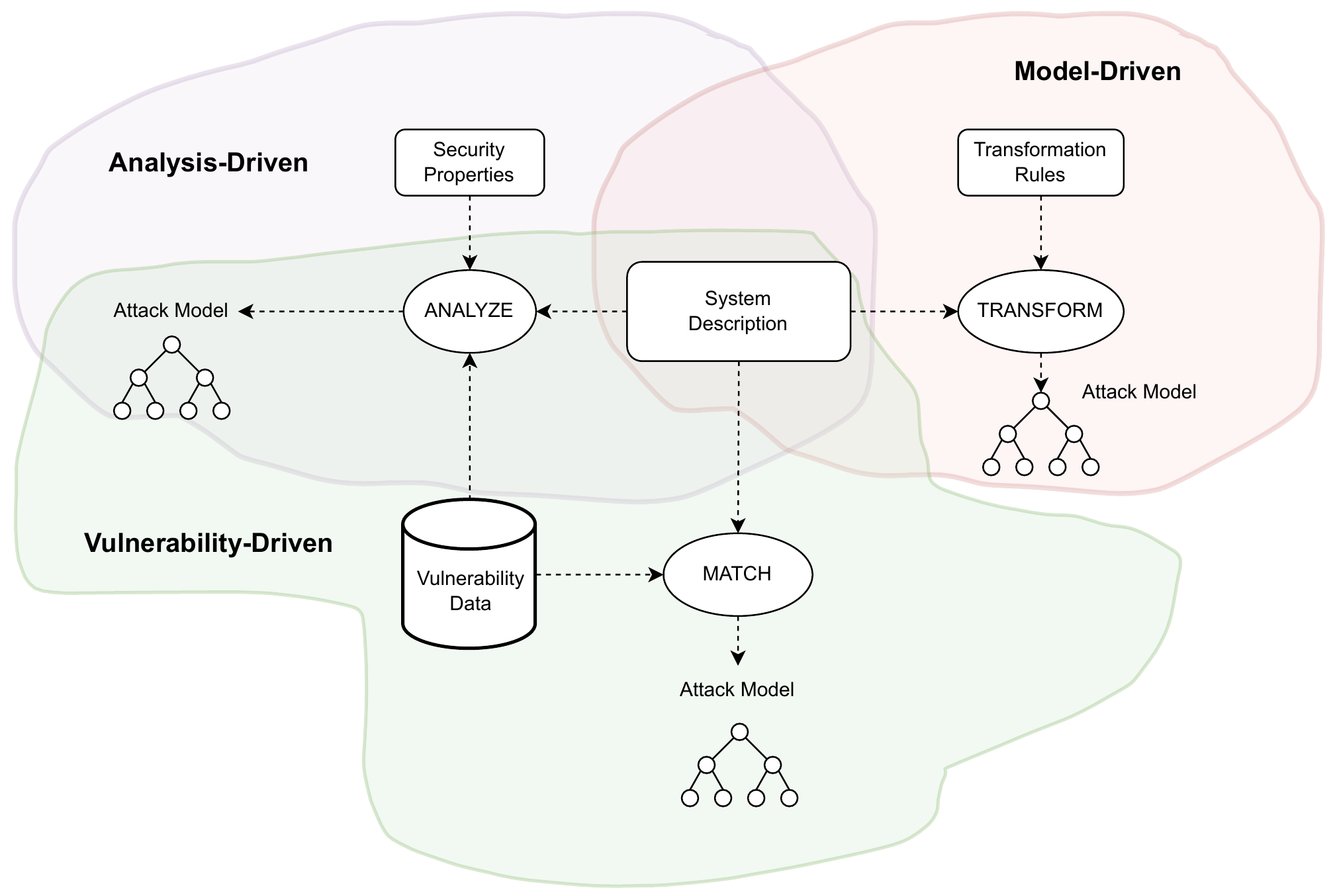}
\caption{Overview of the categories}

\label{fig:overview}
\end{figure*}

In order to provide a structured view of the papers through a common lens, we use the categorization framework visualized in Figure~{\ref{fig:overview}}. The framework has resulted from a careful analysis of all papers with the aim of identifying commonalities among the proposed approaches. We have indeed observed that most papers obtain attack models based on system descriptions or models (Model-Driven approaches), existing vulnerabilities (Vulnerability-Driven approaches), and system analysis results (Analysis-Driven approaches). We will describe these ``categories'' in detail in Section~{\ref{sec:categories}}.


\subsection{Categories} \label{sec:categories}
Figure~{\ref{fig:overview}} provides an overview of the main categories of approaches and their ingredients: Round boxes represent information, while ovals represent processes. The dashed arrows indicate how the processes combine information to produce the attack models. The techniques vary, of course, depending on the specific characteristics of the visualized components. For example, the description of the system can be given in multiple formats, depending on the approach.

As we can see in Figure~{\ref{fig:overview}} all the techniques need some information about the system, but they incorporate additional information (analysis results, known vulnerabilities, etc.) to derive the Attack Model. The three categories (Model-Driven, Analysis-Driven, and Vulnerability-Driven) are not disjoint, and some papers may fall into more than one category, as illustrated in Figure~{\ref{fig:overview}}.

We now describe each category separately.

\begin{itemize}

 \item \textit{Model-Driven:} This class of approaches uses the description of the system as a starting point. Such a description is directly translated into an attack model, typically using some transformation rules.

 \item \textit{Analysis-Driven:} This class of approaches also uses the description of the system as a starting point. Such a description is given as input to a process that analyzes it, for example, checking some specific security property or detecting the occurrence of known vulnerabilities. The results of the analysis are used to synthesize the Attack Model.
 
 \item \textit{Vulnerability-Driven:} This class of approaches starts with collecting relevant data from a vulnerability database or exploiting vulnerability templates or patterns of attack models. The patterns are then matched to the system description in a process that is often iterative.

\end{itemize}

\section{Research Method}\label{sec:3}

In this section, we present the procedure we followed in order to identify the relevant papers to be included in the survey.
%
%
More precisely, we decided to conduct our research using the snowballing technique~\cite{Wohlin}. The snowballing technique refers to the procedure of identifying relevant papers from the reference list or the citations of a selected paper. Our initial set of papers was obtained as described in Section~\ref{sec:start}. The snowballing iterations are described in Section~\ref{sec:iterations}. Our final pool resulted in 20 papers.

\subsection{Start Set} \label{sec:start}
The first step of the snowballing technique is forming the starting set of papers. For this purpose, we had to identify relevant keywords to form a query for a selected database. The keywords we selected to use are: ``Automatic Generation'', ``Automated Generation'', ``Attack Trees'' and ``Attack Graphs''. We performed our research in DTU Findit \\ \url{https://findit.dtu.dk}, which is an open (guest access) database and includes publications from well-known journals and databases. We used the following query: title:(``Automatic generation'' OR ``Automated generation'') AND title:(``Attack trees'' OR ``Attack graphs''). Our search returned 13 papers, of which 3 were not available online. After reading the papers in detail, we discarded five of them for not being relevant. Therefore, we formed our start set, including five papers.
~\cite{Muthumanickam,Zhong,Aksu,Sheyner,Vigo}.

\subsection{Iterations} \label{sec:iterations}
After finding the start set, we have to decide which papers we are going to include in our final pool. For this purpose, we applied \textit{Backward and Forward Snowballing}. Backward snowballing refers to the examination of the reference list of the papers. In order to identify if a paper will be included, we extract some information regarding the title, the author, and the publication venue. Naturally, we should also take into account the context in which the paper is referenced. At this point, if a paper is still in consideration, we read the abstract and other parts of the paper in order to decide if the paper will be included. The forward snowballing is conducted in order to identify papers from the citation list of the paper being examined. Again, for each paper in the citation list, we extracted some basic information, we took into consideration in which context the citation is taking place, and for the final decision, we read parts of the paper~\cite{Wohlin}.

\section{Classification based on the categories}\label{sec:class}


In this section, we present our findings for each category. At the beginning of each category, we present a brief overview of our findings, followed by a detailed presentation of each paper. We also include Tables~\ref{tab:model}-~\ref{tab:vul} to graphically represent the key characteristics of every paper. The first column includes a reference to the paper under examination, and the rest of the columns follow the flow of Figure~\ref{fig:overview}. First, the system description refers to the format of the input system information. For the Model-Driven approaches, the third column summarizes the transformation rules. For the Analysis-Driven, we summarize the security properties and the technology used for the analysis. For the Vulnerability-Driven, the vulnerability data used and which procedure was followed for the instantiate process. Finally, for all the categories, the final column represents the format of the Graph Model produced.



\subsection{Model-Driven Approaches}
 \begin{table*}[]\centering\begingroup\fontsize{6.7pt}{6.7pt}\selectfont    
  \caption{Overview of the papers included in the category \textit{Model-Driven}.  Abbreviated columns are: (A)pproach and M(odel).}
  \label{tab:model}
\begin{tabular}{|c | c | c | c |} 
 \hline
 A & System Description & Transformation Rules &  M  \\ [0.5ex] 
 \hline\hline
 \cite{Gadyatskaya,Ivanova} & \makecell[l]{$\bullet$ A TREsPASS Socio-technical model \\ $\bullet$ A specific asset to be protected \\ $\bullet$ A profile of the attacker} & \makecell[l]{$\bullet$ Feasible ways to gain access to assets} & AT \\
 \hline
   \cite{schiele2021novel}  &   \makecell[l]{$\bullet$ An Attack Tree} &  \makecell[l]{$\bullet$ Tree to graph rules for actions and refinements} & AG \\ 
 [1ex] 
 \hline
\end{tabular}
\endgroup
\end{table*}

All papers falling into this category use a system description as a starting point and then use a set of rules to translate the description into an Attack Model. In comparison with Analysis- and Vulnerability-Driven approaches, the methods in this category do not base the translation on any sophisticated analysis of the security properties of the system or on well-known vulnerabilities. The main reason for this is that the models used as a starting point are already in the form of a threat model (in some cases, one of the two kinds of attack models that we consider).

\paragraph*{\bf Papers~\cite{Gadyatskaya} and ~\cite{Ivanova}} The works by Gadyatskaya~\cite{Gadyatskaya} and Ivanova et al.~\cite{Ivanova} present approaches to Attack-Defense tree generation for TREsPASS socio-technical models~\cite{TRESSPASS}. The approaches are essentially the same, with the main differences being that the work of~\cite{Ivanova} discusses aspects of the mobility of malicious agents in the system, while the work of~\cite{Gadyatskaya} includes a discussion on countermeasures.

\begin{itemize}
\item{\bf System Description.} A socio-technical representation of the system is based on a simplified version of the TREsPASS model~\cite{TRESSPASS}. The model is essentially a graph of locations, assets, and actors, enriched with information about their connections and access control policies.

 \item{\bf Transformation Rules.}
 The transformation rules are based on the conditions (sets of actions) needed for agents to gain access to assets based on the specified connections and locations. This may include gaining access to additional assets.

\item{\bf Transformation and Attack Model Generation.} The rules are applied to the socio-technical model to obtain a set of so-called attack bundles, which represent feasible attacks on individual assets. 
Bundles are then combined to generate one Attack Tree. The main idea is that gaining control of one asset could require gaining control of additional assets. Hence, the corresponding bundles are attached to the leaves. The authors of ~\cite{Gadyatskaya} also discuss how to enrich the model with countermeasures.


\end{itemize}






\paragraph*{\bf Paper~\cite{schiele2021novel}} The authors are proposing a transformation from Attack Trees to Attack Graphs.

\begin{itemize}
 \item{\bf System Description.} An Attack Tree.

 \item{\bf Transformation Rules.} A set of formally defined rules that apply to each case in the tree: leaves representing attack actions and refinements obtained with each attack tree operator (AND, OR), SAND, etc. The rules specify how to transform the corresponding attack tree pattern into the corresponding representation in an attack graph. For example, a leave/action is transformed into an edge with the name of the action on it. 

 \item{\bf Transformation and Attack Model Generation.} The transformation rules are recursively applied to the tree to obtain the corresponding attack graph.

\end{itemize}

 \subsection{Analysis-Driven Approaches}

 \begin{table*}[]\centering\begingroup\fontsize{5.7pt}{5.7pt}\selectfont
    
  \caption{Overview of the papers included in the category \textit{Analysis-Driven}. Abbreviated columns are: (A)pproach and M(odel).}
  \label{tab:analysis}
  
\begin{tabular}{| c | c | c | c | c | c |} 
 \hline
 A & \makecell[c]{System \\ Description} & \makecell[c]{Security  Properties\\ or Vulnerability Data} & \makecell[c]{Analysis} &  M \\ 
 \hline\hline
    \cite{Sheyner,sheyner2003tools} & \makecell[l]{$\bullet$ A NuSMV model of the network \\ \phantom{$\bullet$} and the intruder}  & \makecell[l]{$\bullet$ A temporal logic formula\\$\bullet$ The CVE database} & \makecell[c]{Model \\ Checking} & AG \\
    \hline
    \cite{Ghazo,ibrahim2019automatic} &  \makecell[l]{$\bullet$ Formal security AADL model \\ $\bullet$ A security property} & \makecell[l]{$\bullet$ A temporal logic formula} & \makecell[c]{Model \\ Checking} & AG \\
 \hline
   \cite{bezawada2019agbuilder} & \makecell[l]{$\bullet$ A set of event logs\\$\bullet$ A PDDL model of vulnerabilities} & \makecell[l]{$\bullet$ CVE database} & \makecell[c]{Planning} & AG   \\
  \hline
     \cite{Ingols} & \makecell[l]{$\bullet$ A network map } & \makecell[l]{$\bullet$ A specific resource in the network\\ $\bullet$ 
   Data from Nessus, Sidewinder, Checkpoint\\$\bullet$ CVE \& NVD DBs} & \makecell[c]{Graph\\Reachability}  & AG   \\
   \hline
 \cite{Hong} & \makecell[l]{$\bullet$ A network map} & \makecell[l]{ $\bullet$ Undesired reachability in the network} & \makecell[c]{Graph \\ Reachability} & AT \\
 \hline
 \cite{Vigo} & \makecell[l]{$\bullet$ A Quality Calculus protocol} & \makecell[l]{$\bullet$ Protocol control points to avoid } & \makecell[c]{Static \\ Analysis}  & AT \\ 
 \hline
    \cite{Pinchinat} &  \makecell[l]{$\bullet$ A description of a building\\ $\bullet$ A model of attacker capabilities } & \makecell[l]{$\bullet$ An asset in the building}  & \makecell[c]{Model \\ Checking} & AT \\
     \hline
 \cite{Ou} & \makecell[l]{$\bullet$ A network Configuration\\$\bullet$ Configuration of Machines\\$\bullet$ Attack exploits}&  \makecell[l]{$\bullet$ Logical description of attack} & \makecell[c]{Logical\\Inference} & AG \\
[1ex]
\hline
\end{tabular}
\endgroup
\end{table*}

The papers falling into this category use a process to analyze the input information in order to derive the Attack Model. The input information always includes a description of the system, and on top of that, there might be some additional information needed, such as a specification of security properties or data from a vulnerability database. The information is then analyzed to produce results that are then turned into an attack model.

The papers differ in the kind of language used to describe the system and the underlying model that abstracts part of the system, the kind of security properties for which one wants to identify possible threats, the analysis techniques used, and the kind of attack model generated. Most papers use a sort of feasibility analysis to extract possible ways of enacting a threat, which is then summarized in the attack model. The kind of analysis can range from simple reachability in a graph to more sophisticated analysis based on model checking, constraint satisfaction, or planning, where concrete threats are specified using some suitable language, typically based on logic.

\paragraph*{\bf Papers~\cite{Sheyner,sheyner2003tools}}
Sheyner et al.~\cite{Sheyner,sheyner2003tools} present a tool-supported approach that generates Attack Graphs for given computer networks using the model checker nuSVM~\cite{nuSVM}.

\begin{itemize}
 \item{\bf System Description.} The system is described as an NuSVM model that specifies several components: a set of hosts, a connectivity relationship, a trust relationship among the hosts, an intruder model, a set of individual actions the intruder can exploit to design an attack, and an intrusion detection system.

 \item{\bf Security Properties.} Security properties are specified as NuSMV temporal properties and data from the Common Vulnerabilities and Exposures (CVE) database.

 \item{\bf Analysis and Attack Model Generation.} The model checker takes as input the property and the model M (the network). If the property is satisfied, the model checker returns true; otherwise, it provides a counterexample. The counterexample depicts a path that the attacker can follow to violate the security property. The paths generated through the analysis step, are then combined to generate the Attack Graph.

\end{itemize}

\paragraph*{\bf Papers ~\cite{Ghazo,ibrahim2019automatic}} Ibrahim et al.~\cite{Ghazo,ibrahim2019automatic} present a model-checking based approach for generating Attack Graphs from AADL~\cite{sei} architectural models of computer and control SCADA networks.

\begin{itemize}
 \item{\bf System Description.} The description of the system's architecture is specified in the AADL Language~\cite{sei}.

 \item{\bf Security Properties.} Security properties are specified with CTL temporal logic formulas expressing the compromise of an asset in the architecture.

 \item{\bf Analysis and Attack Model Generation.} The analysis takes as input the model of the architecture and the security property. The mode checker JKind~\url{http://loonwerks.com/tools/jkind.html} is used to check security properties and to automatically produce counterexamples. Counterexamples are then used to generate the Attack Graph.

\end{itemize}

\paragraph*{\bf Paper~\cite{bezawada2019agbuilder}} Bezawanda et al.~\cite{bezawada2019agbuilder} present a planning-based tool that automatically generates Attack Graphs from vulnerability databases.

\begin{itemize}
 \item{\bf System Description.} The system description consists of a set of event logs and a PDDL~\cite{ghosh2012planner} specification of a planning domain and problem.
%
%
 The approach uses natural language processing to produce the PDDL specification from the textual description of vulnerabilities from the Common Vulnerabilities and Exposures(CVE)~\url{https://cve.mitre.org/} or the National Vulnerability Database (NVD)~\url{https://nvd.nist.gov/}. 

 \item{\bf Security Properties.} The security property is specified as a goal in the PDDL planning problem. 

 \item{\bf Analysis and Attack Model Generation.} The event logs of the system are used to bind the abstract variables of the PPDL planning domain to form a concrete PPDL problem for the specific system. The PDDL problem is then solved with the help of the PDDL planner, which tries to find a plan, i.e., a sequence of actions the attacker must perform to achieve its goal. Planning solutions are then combined to produce an Attack Graph. 

\end{itemize}

\paragraph*{\bf Paper~\cite{Ingols}} Ingols et al.~\cite{Ingols} present an approach to generate attack graphs for computer network vulnerabilities.

\begin{itemize}
 \item{\bf System Description.} The system description is in the form of a map of the computer network under study.

 \item{\bf Security Properties.} The security properties under study are the reachability of specific resources in the newtork. 

 \item{\bf Analysis and Attack Model Generation.} The analysis computes a reachability matrix for the network, which indicates how network resources can be reached from each other. The analysis also imports data from Nessus \url{http://www.nessus.org}, Sidewinder, and Checkpoint firewalls, as well as from the CVE repositories\url{http://cve.mitre.org} and NVD \url{http://nvd.nist.gov.}. From such data and from the network,
 the analysis generates multiple so-called prerequisite graphs, which consist of three different kinds of nodes: state nodes, prerequisite nodes, and vulnerability instance nodes.
\end{itemize}

\paragraph*{\bf Paper~\cite{Hong}} Hong et al.~\cite{Hong} present an approach to summarize all attack paths in a computer network as an Attack Tree.

\begin{itemize}
 \item{\bf System Description.} The system description is in the form of a specification of the computer network under study.

 \item{\bf Security Properties.} A specific node in the network to be reached by the attacker.

 \item{\bf Analysis and Attack Model Generation.} The analysis computes all possible ways for an attacker to reach the target node in the network. Such paths are then combined into an Attack Tree. After the construction of the full Attack Tree, which users can shrink or expand based on a selection of suitable parameters.

\end{itemize}

\paragraph*{\bf Paper \cite{Vigo}}
Vigo et al.~\cite{Vigo} propose a technique to extract attack trees from protocols used in distributed systems specified in the Quality Calculus~\cite{QC} language using static analysis.

\begin{itemize}
\item{\bf System Description.} The system description is assumed to be given using the Quality Calculus~\cite{QC} (an extension of the $\pi$-calculus).
Such a description specifies how several processes interact via message-passing and how they process the information received using pattern matching on data.

\item{\bf Security Properties.} The main security property provided as input is the reachability of a specific program point \textit{l} in the protocol, typically an unfortunate state that one wishes to avoid.

\item{\bf Analysis and Attack Model Generation.}
The main idea of the analysis is to determine the possible ways to access a specific program point \textit{l}. The backward chaining exposes all the paths leading to point \textit{l}, thus unveiling all the conditions that the data being sent must satisfy in order to reach that \textit{l}.
The paths leading to the program point \textit{l} are then combined in an Attack Tree.

\end{itemize}

\paragraph*{\bf Paper~\cite{Pinchinat}} Pinchinat et al.~\cite{Pinchinat} present ATSyRA, an Eclipse tool available at \url{https://atsyra2.irisa.fr/} to generate Attack Trees for military buildings.

\begin{itemize}
 \item{\bf System Description.} A description of a building and of the attacker's capabilities, is specified as a formal model.

\item{\bf Security Properties.} An attack objective, i.e. an asset of the building to be compromised by the intruder.

 \item{\bf Analysis and Attack Model Generation.} The analysis is based on a model checking the property of compromising the asset and generating an Attack Graph as an intermediary step, which specifies the possible paths of actions needed to reach the asset. Based on the Attack Graph, the security expert specifies a set of high-level actions (HLA), which describe how each attacker action in the graph can be refined into sub-actions. The final step of the analysis uses the HLA actions and the attack graph to construct the Attack Tree.

\end{itemize}

\paragraph*{\bf Paper~\cite{Ou}}
Ou et al.~\cite{Ou} present an approach based on MulVAL~\cite{ou2005mulval}, a system for automatically identifying security vulnerabilities in networks. The central notion of MulVAL is that the configurations of a network and its machines are represented as Datalog tuples and security semantics and attacks are represented as Datalog rules.

\begin{itemize}
 \item{\bf System Description.}
 The system description is in the form of a Datalog specification of the network configuration, the configuration of machines, the semantics of security aspects of machines, and the description of attack exploits.

\item{\bf Security Properties.} The security properties for which one wants to obtain an attack model are specified with logical rules.

\item{\bf Analysis and Attack Model Generation.}
The proposed method then analyzes the input by evaluating the Datalog rules, using the MulVal reasoning engine, which produces a set of traces realizing the attacks.
The traces are sent to a graph-building algorithm, which combines them to produce an Attack Graph. 

\end{itemize}

\subsection{Vulnerability-Driven Approaches}
\begin{center}
 \begin{table*}[]\centering\begingroup\fontsize{5.7pt}{5.7pt}\selectfont
    
  \caption{Overview of the papers included in the category \textit{Vulnerability-Driven}.  Abbreviated columns are: (A)pproach and M(odel). }
  \label{tab:vul}

\begin{tabular}{|c | c | c | c | c | c |} 
 \hline
 Paper & \makecell[c]{System \\ Description} & \makecell[c]{Vulnerability Data} & \makecell[c]{Instantiate\\Match} &  M  \\
 \hline\hline
   \cite{Phillips}  & \makecell[l]{$\bullet$ A set of templates \\ $\bullet$ A configuration of the system} & \makecell[l]{$\bullet$ A library of Attack-Tree templates\\$\bullet$ An attacker profile} & \makecell[c]{Pattern\\Matching} & AG   \\
 \hline
 \cite{Swiler} &    \makecell[l]{$\bullet$ A configuration of the network }\ & \makecell[l]{$\bullet$ A library of Attack-Tree templates\\ $\bullet$ An attacker profile} & \makecell[c]{Pattern\\Matching} & AG  \\
 \hline
   \cite{Tippenhauer}  &  \makecell[l]{$\bullet$ A workflow of the system \\$\bullet$ A security goal }l&  \makecell[l]{$\bullet$ A library of Attack-Tree templates\\ $\bullet$ An attacker mode} & \makecell[c]{Pattern\\Matching} & AG  \\
 \hline
  \cite{koo2022attack}  & \makecell[l]{$\bullet$ A network topology  \\ $\bullet$ Network information} & \makecell[l]{$\bullet$ A trained model for recognizing \\\phantom{$\bullet$} attack paths} &  \makecell[c]{Deep\\Learning} & AG  \\
 \hline
  \cite{Bryans} &  \makecell[l]{$\bullet$ A computer network map}& \makecell[l]{$\bullet$ A library of Attack-Tree templates}& \makecell[c]{Pattern\\Matching} & AT \\ 
 \hline
 \cite{Kumar}  & \makecell[l]{$\bullet$ A generic system description} & \makecell[l]{$\bullet$ An Attack-Tree template} & \makecell[c]{Pattern\\Matching} & AT  \\
 \hline
  \cite{Pinchinat_2} & \makecell[l]{$\bullet$ An event log} & \makecell[l]{$\bullet$ A CFG of valid attack traces} & Parsing & AT  \\
  \hline
 \cite{jhawar2018semi} & \makecell[l]{$\bullet$ An initial version of Attack Tree}   & \makecell[l]{$\bullet$ A library of annotated attack trees\\\phantom{$\bullet$\ }based on NDV and CAPEC}  & Library & AT  \\
 \hline
  \cite{gadyatskaya2017refinement} &  \makecell[l]{$\bullet$ A system model \\ $\bullet$ A refinement specification \\ $\bullet$ Intended semantics}& \makecell[l]{$\bullet$ A trace of successful attacks} & Successful Attacks  & AT   \\ 
 \hline
\end{tabular}
\endgroup
\end{table*}
\end{center}

All the papers falling into this category use a process to match some vulnerability patterns to information about the system. In some cases, the system is assumed to be specified in detail (e.g., as the configuration of a computer network), while others assume that the only available information is system logs or human experts that can be queried. The patterns are often Attack Tree or Graph fragments that can be used to refine or extend a current Attack Model. Some approaches are semi-automatic and need human intervention to decide how to instantiate the patterns and how to proceed with the Attack Model refinement, while others are fully automated.

In the following summary of the papers, we have excluded those papers that also fall into the Analysis-Driven category and that use vulnerability data (namely~\cite{bezawada2019agbuilder,Ingols,Sheyner,sheyner2003tools}) to avoid redundancies, but they too fall into this category.

\paragraph*{\bf Paper~\cite{Phillips}} Phillips et al.~\cite{Phillips} present an approach to generate Attack Graphs for computer networks.

\begin{itemize}
 \item{\bf System Description.} The system description is given in the form of the configuration of a computer network.

 \item{\bf Vulnerability Data.} Vulnerability data is composed of a set of attack graph templates capturing common attacks along with the attacker profiles.

 \item{\bf Match and Attack Model Generation.} The procedure starts at the target node for the attacker and iteratively tries to match and instantiate the attack templates. The procedure is repeated until a node in the network is reached, which is considered to be accessible to the attacker.

\end{itemize}

\paragraph*{\bf Paper~\cite{Swiler}} Swiler et al~\cite{Swiler} propose a method to generate an attack graph for a computer network.

\begin{itemize}
 \item{\bf System Description.} The configuration of the computer network under study.

\item{\bf Vulnerability Data.} Vulnerability data comes in the form of attack graph templates and attacker profiles. The attack templates represent steps of known attacks or strategies for moving from one state to another in a network.

\item{\bf Match and Attack Model Generation.} The tool instantiates the attack templates according to the attacker profile and the network configuration.
 At each instantiation step, when a node in a graph matches the requirements of the template, a new edge is added to the network and new nodes are created. When no more templates can be matched, the process results in an attack graph.

\end{itemize}

\paragraph*{\bf Paper~\cite{Tippenhauer}} Tippenhauer et al.~\cite{Tippenhauer} present a work to generate attack graphs for workflows.

\begin{itemize}
 \item{\bf System Description.} The system description is mainly a workflow for the system. 

 \item{\bf Vulnerability Data.} Vulnerability data is based on attack graph templates and attacker models. Templates are generated from the security goal and the workflow.


\item{\bf Match and Attack Model Generation.}
 The approach starts with an attack graph based on a specific security goal for the attacker. The procedure then extends the initial graph every time a template can be matched with the given workflow description.

\end{itemize}

\paragraph*{\bf Paper~\cite{koo2022attack}} Koo et al.~\cite{koo2022attack} introduce a method to support the generation of a Attack Graphs for computer networks using machine learning.

\begin{itemize}
 \item{\bf System Description.} The system description is given in the form of the network topology and additional system information.

\item{\bf Vulnerability Data.} The approach uses machine learning to train a model to recognize attack paths in attack graphs. The training data is based on vulnerability databases.

 \item{\bf Match and Attack Model Generation.}
 The process of generating an attack graph consists of using the trained model to recognize attack graph features in the given network topology.

\end{itemize}

\paragraph*{\bf Paper~\cite{Bryans}} Bryans et al.~\cite{Bryans} introduce a method to generate Attack Trees automatically from the description of a computer network and a set of templates.

\begin{itemize}
 \item{\bf System Description.} The description of the computer network under study

 \item{\bf Vulnerability Data.} The vulnerability data is given in the form of a set of attack tree templates that represent possible attacks.

\item{\bf Match and Attack Model Generation.} Templates are matched against the network by instantiating the variables in the template with the actual components of the computer network under investigation. The leaves in the resulting matched template can be refined using the same procedure, yielding a complex attack tree. The tool is available at \url{https://tinyurl.com/uoptgfb}.

\end{itemize}

\paragraph*{\bf Paper~\cite{Kumar}} Kumar et al.~\cite{Kumar} introduce a library of attack templates that can be used to refine and instantiate for a specific system.

\begin{itemize}
\item{\bf System Description.} No specific system description format is required.

 \item{\bf Vulnerability Data.} The approach proposes a set of layered Attack Tree templates that have been constructed based on the literature on Attack Trees to integrate their common characteristics.

\item{\bf Match and Attack Model Generation.} As for several of the approaches discussed in this section, the approach works by refining the templates iteratively based on specific characteristics of the system under study.

\end{itemize}

\paragraph*{\bf Paper~\cite{Pinchinat_2}} Pinchinat et al.~\cite{Pinchinat_2} propose a tool (\url{http://attacktreesynthesis.irisa.fr/}) to build Attack Trees from system logs.

\begin{itemize}
 \item{\bf System Description.} No actual system description is used, apart from a set of attack traces of the system (logs of events that resulted in a vulnerability). 

 \item{\bf Vulnerability Data.} Vulnerability data is given as context-free grammar that represents valid threat traces.

 \item{\bf Match and Attack Model Generation.}
 The context-free grammar is used to parse the trace log and generate an attack tree akin to a parse tree.

\end{itemize}

\paragraph*{\bf Paper~\cite{jhawar2018semi}} Jhawar et al.~\cite{jhawar2018semi} present a semi-automatic procedure for creating Attack Trees based on an initially given proposal and alternating manual manipulations with automatic refinements.

\begin{itemize}
 \item{\bf System Description.} The system description consists of an initial Attack Tree that describes some tentative threats to the actual system.

\item{\bf Vulnerability Data.} Vulnerability data consists of a library of annotated attack trees based on the National Vulnerability Database~\url{https://nvd.nist.gov/} and the Common Attack Pattern Enumeration and Classification~\url{https://capec.mitre.org/}.

 \item{\bf Match and Attack Model Generation.} The automated steps of the process to refine a tree consist of finding instances of the templates in the current tree proposal. Trees are annotated with predicates that determine if an Attack Tree can be attached to another attack tree as a sub-tree. The final Attack Tree is obtained when the manual and automated steps yield a result that is satisfactory for the users.

\end{itemize}

\paragraph*{\bf Paper~\cite{gadyatskaya2017refinement}} The authors in~\cite{gadyatskaya2017refinement} propose a method to generate Attack Trees from successful attacks.

\begin{itemize}
 \item{\bf System Description.} The system description is actually not provided. Instead, one departs from a set of logs of malicious activities.

 \item{\bf Vulnerability Data.} Vulnerability data is given by a set of valid refinements of Attack Trees.

 \item{\bf Transformation and Attack Model Generation.}
 A greedy heuristic for encoding and decomposing attack traces is used to generate the Attack Tree, according to the specification of valid refinements.

\end{itemize}

\section{Limitations, Challenges and Future Directions}\label{sec:lim}
We start describing in Section~{\ref{sec:dimensions}} some features called \emph{dimensions} that we used to identify gaps between the existing approaches.
We use those dimensions to analyze the surveyed papers in Section~{\ref{sec:limitations}}).
Based on our analysis we discuss a set of research challenges and future directions in  Section~{\ref{sec:future}}). 
Our aim is to serve as inspiration for future directions for the research community. 
%
Subsequent sections discuss possible connections with other models similar or extending Attack Trees and Attack Graphs, namely fault trees (Section~{\ref{sec:FT}}), attack-defence trees (Section~{\ref{sec:ADT}}), and attack-defence diagrams (Section~{\ref{sec:ADD})}.

\subsection{Dimensions} \label{sec:dimensions}

Orthogonally, our study aims at spotting the gap between the research on this field and its actual adoption in practice and helping the reader identify which technique suits their case best. Hence, having these two points in mind, we selected some features worth studying in each paper in order to address the aforementioned consideration. We call these features ``dimensions'':

\begin{itemize}
 \item \textit{Proofs:} Whether a paper includes formal proofs for the algorithms applied for generating the Attack Models. A proper formalization is important in order to have rigorous guarantees about the use of the approach.
 \item \textit{Open-Source:} Whether a paper includes a reference to the implementation of the proposed solution and if the code is available online in an open-source repository. This dimension is important for prospective users to understand if the proposed approach is supported by tools that can be used immediately or if they will have to obtain the tools from the authors or, in the worst case, re-implement the tools based on the paper description. This dimension is related to ``reproducibility'' of the results (discussed below).

\item \textit{Reproducibility:} Whether the authors conducted experiments to validate their approach and if the paper provides detail to reproduce those experiments, witnessed by a reproducibility badge. This dimension is important to assess the scientific validity of the experiments and for prospective users to assess how the methods work on their own systems. We observed that none of the papers have a reproducibility badge.

 \item \textit{Scalability:} Whether the authors discuss the scalability of their solution to show that it can be applied in real-life scenarios.
\end{itemize}

\begin{table*}[]\centering\begingroup
\fontsize{9.0pt}{9.0pt}\selectfont
    
  \caption{Overview of dimensions and limitations. Abbreviated columns are: (A)pproach, M(odel), (R)eproducibility, (O)pen-source, (P)roof and (S)calability.}
  \label{tab:dim}
\smallskip
\begin{tabular}{| c | c | c | c | c |} 
 \hline
 A & R & O & P & S  \\ [0.5ex] 
 \hline\hline
   \cite{Phillips}   &  & &  & \\
 \hline
   \cite{Sheyner,sheyner2003tools} & \xmark &  & \checkmark & \\
 \hline
\cite{Ghazo,ibrahim2019automatic} & \xmark &   & \checkmark  & \checkmark \\
 \hline
 \cite{Swiler}   &  & &  & \\
 \hline
   \cite{Tippenhauer}   & &  & \checkmark & \\
 \hline
 \cite{koo2022attack}   & \xmark & &  & \\
 \hline
  \cite{bezawada2019agbuilder}  &  & &  &  \\ 
  \hline
   \cite{Ingols}  & \xmark  &\  &  & \checkmark \\
  \hline
 \cite{Bryans}  &\xmark & \checkmark &  & \\ 
 \hline
 \cite{Kumar}  & &  & & \\
 \hline
  \cite{Pinchinat_2} &  & \xmark  & \checkmark & \checkmark\\
  \hline
\cite{jhawar2018semi}  &   &\checkmark & \checkmark & \\
 \hline
 \cite{Hong} & \xmark  & &  & \checkmark\\
\hline
 \cite{Vigo} & &  & \checkmark & \checkmark \\ 
 \hline
 \cite{Gadyatskaya,Ivanova}  & &  &  & \\
 \hline
   \cite{schiele2021novel}  & &  & \checkmark & \\
   \hline

 \cite{Ou} & \xmark &  & \checkmark & \checkmark\\
 \hline

   \cite{Pinchinat} &  & \xmark  &  & \\
 \hline
   \cite{gadyatskaya2017refinement} &  &   & \checkmark & \\
  \hline

\end{tabular}
\endgroup
\end{table*}

\subsection{Analysis of Dimensions and Limitations} \label{sec:limitations}

In this subsection, we provide an overview of the limitations we found in the surveyed papers. The summary is provided in Table~{\ref{tab:dim}}.
We denote the symbol \checkmark~when the referred paper fulfills the corresponding characteristics for Reproducibility, Open-source, Proof, and Scalability. For Reproducibility, \xmark~denotes that experiments are provided, but no reproducibility badge is exhibited. For Open-source, \xmark~denotes that a tool is available, but \checkmark~indicates that it is also an open-source project.

The main limitations we have identified are:

\begin{itemize}
\item All approaches from the model-driven category depart from an initial threat model. Hence, users must have had a preliminary threat modeling analysis to exploit the approach to derive an Attack Model, which in part defeats one of the main goals of automated Attack Model generation approaches, namely to provide an initial threat model. 

\item Analysis-driven approaches often suffer from the complexity of the analysis procedures. For example, model-checking approaches suffer from the state-space explosion, which may render these approaches unsuitable for real-life cases. 

\item Vulnerability-driven approaches often use static templates to form the Attack Models. Static templates, cannot be updated automatically, so the user should constantly update the library of templates. Templates aim to serve all kinds of systems, and this can lead to generic models. Also, the limited number of templates can produce incomplete Attack Models. Some other approaches in this category use traces of past attacks to extract vulnerabilities. This is specified in the corresponding system, but the logs are not always accessible. Last, templates are often generic and may not apply to specific novel systems.

\end{itemize}

A generic limitation of all approaches is that automatically obtained models may not be meaningful for the users, which ultimately requires users to play an active role in the refinement of the models.

\subsection{Research Challenges and Future Directions} \label{sec:future}
\begin{itemize}
 \item \textit{Attack Tree Extensions:} In the automatic generation of Attack Trees most papers focus on basic attack trees with $\mathit{AND}$ and $\mathit{OR}$ operators. Only a few papers are including the $\mathit{SAND}$ operator (see Section~\ref{sec:2}. The $\mathit{SAND}$ operator can represent multiple situations that occur in different kinds of attacks~\cite{Jhawar}. So we believe that it is important to also include this operator in the automatic generation. Likewise, we believe it would be interesting to explore richer models of attack trees with additional operators proposed in the literature, such as $k-out-of-N$ and defensive measures (see detailed discussion of attack-defense trees (Section~\ref{sec:ADT}) and attack-defense diagrams (Section~\ref{sec:ADD})). 
\item \textit{Dynamic Solutions:} Systems are being upgraded constantly. New variables or configurations can make an Attack Model useless. The need for more dynamic solutions is crucial. Whenever there are some changes in the system, the current version of Attack Models should be updated, helping security experts identify new potential threats. As far as we know, there is only one paper in the literature proposing a more dynamic solution~\cite{Phillips}.
 \item \textit{Lack of Formalization:} The use of semantics and proofs is really important. There are some works that do not use proofs or semantics to support their solution. Neglecting semantics can lead to poor or meaningless results. The use of proofs can help the reader better understand the concepts used and help the researchers expand the field. Semantics can help the community develop the same language concerning the generation of Attack Models.
 \item \textit{Forensics:} Another field important to security experts is forensics. Forensics can help security experts identify vulnerabilities that were not previously taken into account and secure their systems better. Using forensics, one can generate Attack Models to depict what goes wrong in a system. This procedure requires the logs or traces of a running system.
\item \textit{Meaningful Results:} The automatic generation of Attack Models provides a very good tool for security experts. But it might produce meaningless results that are not usable. There are not many works providing proper means to prevent the generation of meaningless results. Mostly semi-automated procedures have been proposed to avoid the generation of meaningless results~\cite{jhawar2018semi, Pinchinat_2, Kumar}. Another work that considers it is~\cite{gadyatskaya2017refinement}.
 \item \textit{Prerequisites:} All of the papers require some prerequisites in order to generate the Attack Models. In some cases, this might be tedious for the security experts if the prerequisites require their involvement (semi-automated), but a fully automated procedure might produce meaningless results. Some prerequisites can be the configuration of the system, which is a very important file that an attacker can exploit to take control of the system. So the prerequisites can themselves be a vulnerability. It is important to find the right balance between the number of prerequisites and producing meaningful results. It is also important to examine the nature of the prerequisites and how crucial they can be for the security of the system.
 \item \textit{Scalability:} Scalability is one of the main characteristics of a tool. Nowadays, we use very large systems to cover our needs. The automatic generation of Attack Trees should be adapted to perform under these circumstances.
 \item \textit{Attack Defence Trees:} A few papers are investigating the automatic generation of Attack Defence Trees~\cite{Gadyatskaya}. After the automatic generation of Attack Models, is it natural to investigate the automatic generation of the corresponding defences/countermeasures.
 \item \textit{Attack Graphs $\leftrightarrow$ Attack Trees}: It would be interesting to explore the transformation of Attack Graphs to Attack Trees and vice versa. Pinchinat et al. are exploiting an Attack Graph to produce an Attack Tree~\cite{Pinchinat}. Also, in~\cite{schiele2021novel} the authors propose a transformation of an Attack Tree to an Attack Graph. This method can be explored further and tested to produce more concrete results regarding this transformation. Further experiments can be performed, especially on the results of automatic generation techniques.
\end{itemize}

\subsection{Fault Trees} \label{sec:FT}

Fault Trees are graphical models that represent how a component failure can lead to a system failure. They were first introduced in the 1960s at Bell Labs. They are Directed-Acyclic Graphs including leaves and gates. The leaves represent component failures and the gates through which these failures can propagate through the system. The most common gates used in Fault Trees are:
\begin{itemize}
\item \textbf{AND gate:} The output event occurs only if all input events occur.
\item \textbf{OR gate:} The output event occurs if at least one of the input events occurs.
\item \textbf{k/N gate:} The output event occurs if at least k out of N input events occur.
\item \textbf{Inhibit gate:} The output event occurs if the input event occurs, and the condition drawn to the right of the gate also occurs. It is like an AND gate with two inputs~\cite{RUIJTERS201529}.
\end{itemize}

One can see the similarities between Fault Trees and Attack Trees. They both represent the basic event of interest at the top (the root of the tree), which is then refined into actions until basic actions can no longer be refined. These models have many similarities, but some extensions in each model lead them to grow in different directions~\cite{ft}. Despite the fact that Fault Trees are out of the scope of this paper, we believe that the similarities between these two models can be exploited in order to explore the different techniques used for the automatic generation of Fault Trees~\footnote{Several authors cite this survey~\cite{berres2016automatic}, which we have not been able to obtain from the authors or the publisher} and their adoption in Attack Trees.

\begin{figure}[t!]
\centering
\includegraphics[width=0.6\textwidth]{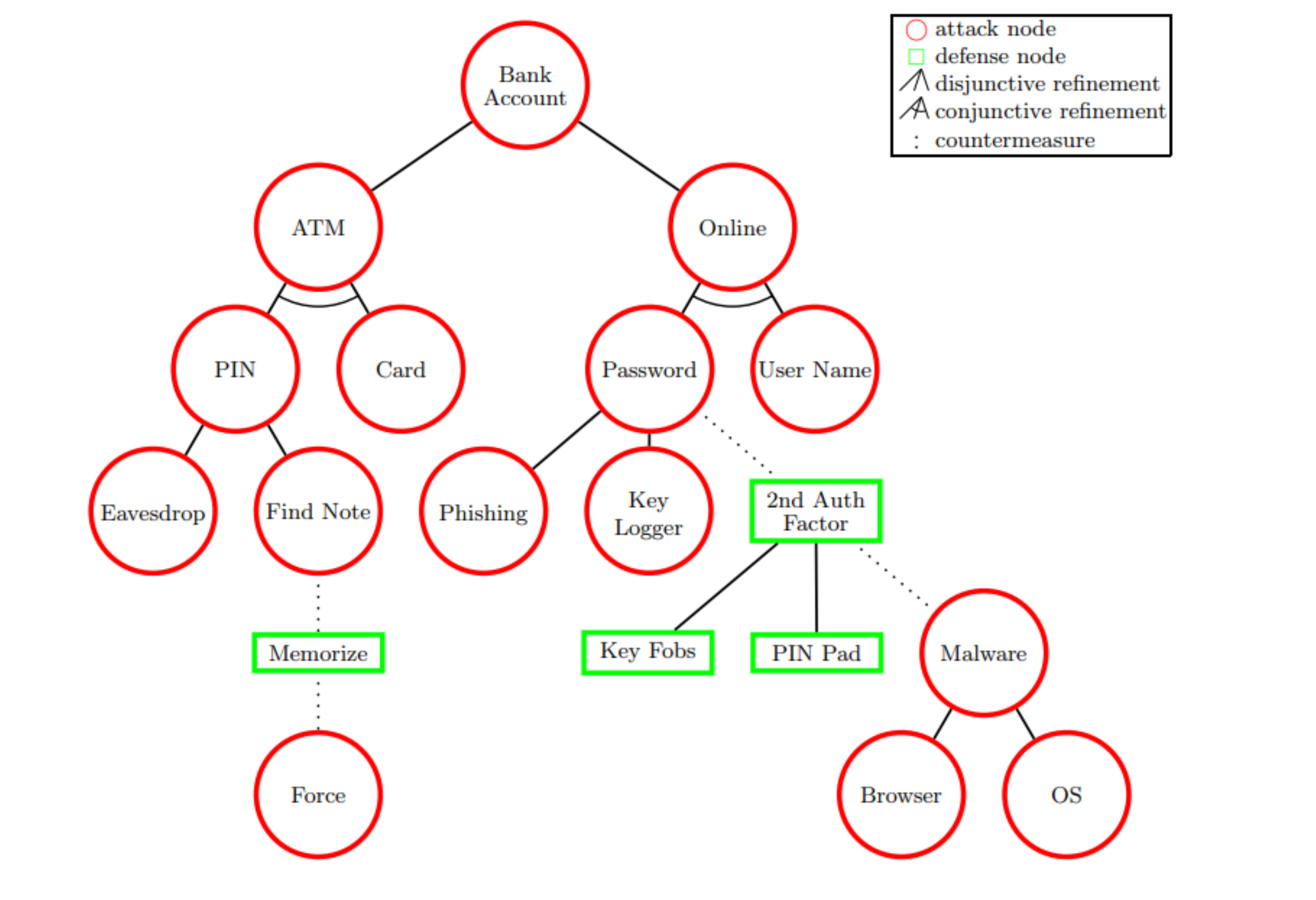}
\caption{Attack-Defense Tree form ~\cite{adt}}
\label{fig:adt}
\end{figure}

\subsection{Attack-Defense Trees} \label{sec:ADT}

Attack-Defense Trees are an extension of the Attack Trees. They consist of a root node, a set of actions an attacker might take to attack a system, and a set of defense mechanisms the defender can employ. On the Attack-Defense Trees, we have two kinds of nodes, the attack nodes, and the defend nodes, that both can be refined into sub-goals. Also, every node can have a child of the opposite type, called a countermeasure. One example of an Attack-Tree can be seen in Figure~\ref{fig:adt}~\cite{adt, kordy2014attack}.

We can observe in Figure~\ref{fig:adt} that the Attack-Defense Tree uses the same operators as the Attack-Tree. The example depicts a bank account attack. The attack can be achieved either from an ATM or online. We can see in the green rectangles the proposed defenses the defender can employ.

In the scope of this paper, we are not going to examine specifically the automatic generation of Attack-Defense Trees. It is an upcoming field, and it is included in the future directions proposed at the end of the paper. We strongly believe that the automatic generation of Attack-Defense Trees can be the next step in the automatic generation of Attack-Trees/Graphs. We are listing some works focused on the automatic generation of Attack-Defense Trees~\cite{Gadyatskaya, moh, Salva2019ACA, eck, siu,sal}.

\subsection{Attack-Defence Diagrams} \label{sec:ADD}

Attack-Defence Diagrams are acyclic graphs that represent how basic actions (events) can be combined to form a game between the attacker and the defender. The game can be either undecided, a successful attack, or a successful defense. This structure is considered to be more expressive than Attack-Defence Trees. The outcome of the game depends, among others, on parameters like time, cost, and probabilities. These parameters decorate the basic event on the Attack-Defence Diagram. This structure is equipped with more expressive gates than Attack-Defence Trees, including the possibility of repetitive behavior~\cite{add}. For further information about Attack-Defense Diagram, we encourage the reader to refer to~\cite{add}. Despite the fact that the Attack-Defense Diagrams are out of the scope of this paper, to our knowledge, there is no available research on how to automatically generate Attack-Defense Diagrams. We believe that it would be an asset to the scientific community to further investigate this model, as it offers more expressive ways to represent an attack-defense scenario.



\section{Conclusion}\label{sec:con}
%

We have provided a survey of the state-of-the-art techniques for the automatic generation of Attack Trees and Attack graphs, driven by three main research questions:
\begin{itemize}
 \item RQ1 What kinds of techniques have been proposed for automatically generating Attack Models?
 We answered this question by providing a comprehensive and structured literature review of the field. Our survey provides a classification based on the procedure followed to derive the Attack Model. We identified three main categories: Model-Driven approaches, where an input model of the system under study is translated to an Attack Model; Analysis-Driven approaches, where a process analyzes and combines the description of the system with additional information; and Vulnerability-Driven approaches, where existing vulnerabilities are adjusted to the system under study.
 Based on these categories, our survey describes the specifics of each proposed approach under the same lens, which helps in relating them.
 .

 \item RQ2 - What are the characteristics of the proposed approaches in terms of formal rigor, empirical evidence, reproducibility, scalability, etc.?
 We answered this question by identifying key dimensions in Section~{\ref{sec:cat}} and presenting an analysis of the papers based on the dimensions in Table~{\ref{tab:dim}}. We consider these dimensions to be important to identify limitations and also to help the reader choose a suitable method to apply.

 \item RQ3 - What are the challenges and opportunities in this area of research? We answer this question in Section~{\ref{sec:lim}} where we discuss the limitations and future directions for the research community. We believe that there is a gap between research and the real employment of the methods, based on our findings about some components, such as scalability and meaningful results. We provide some future directions to the research community to eliminate such a gap.
\end{itemize}

Taking everything into consideration, we believe that our work has a twofold result: (1) it helps the reader navigate through the literature and decide on the methods and technologies that are more suitable for them, and (2) it provides limitations and future directions for the scientific community.

\section*{Acknowledgment}

This work has been supported by Innovation Fund Denmark and the Digital Research Centre Denmark, through the bridge project ``SIOT – Secure Internet of Things – Risk analysis in design and operation''.

\bibliographystyle{plain}
\balance
\bibliography{bibliography}

\end{document}